\documentclass[conference]{IEEEtran}
\IEEEoverridecommandlockouts
\usepackage{cite}
\usepackage{amsmath,amssymb,amsfonts}
\usepackage{algorithmic}
\usepackage{subcaption}
\usepackage{graphicx}
\usepackage{url}
\usepackage{booktabs}
\usepackage{textcomp}
\usepackage[colorinlistoftodos]{todonotes} 
\usepackage{xcolor}
\def\BibTeX{{\rm B\kern-.05em{\sc i\kern-.025em b}\kern-.08em
    T\kern-.1667em\lower.7ex\hbox{E}\kern-.125emX}}

\usepackage{fancyhdr}
\makeatletter
\let\old@headrule\headrule
\renewcommand{\headrule}{\if@fancyplain\let\headrulewidth\plainheadrulewidth\fi\old@headrule}
\makeatother

\renewcommand{\headrulewidth}{0pt}

\makeatletter
\def\ps@IEEEtitlepagestyle{%
  \def\@oddhead{\mbox{}\scriptsize\rightmark \hfil}%
  \def\@evenhead{\scriptsize\thepage \hfil \leftmark\mbox{}}%
  \def\@oddfoot{\hfil \mbox{}\parbox{5.5in}{\centering
  \footnotesize \textcopyright 2025 IEEE. Personal use of this material is permitted.
  Permission from IEEE must be obtained for all other uses, in any current or future
  media, including reprinting/republishing this material for advertising or promotional
  purposes, creating new collective works, for resale or redistribution to servers or
  lists, or reuse of any copyrighted component of this work in other works.}\hfil \mbox{}}%
  \def\@evenfoot{\mbox{}\parbox{5.5in}{\centering}\hfil \mbox{}\hfil}%
}
\makeatother
\IEEEoverridecommandlockouts

\begin{document}

\newcommand{\footnoteref}[1]{\textsuperscript{\ref{#1}}}

\title{Deceptive Game Design? Investigating the Impact of Visual Card Style on Player Perception}

\author{\IEEEauthorblockN{1\textsuperscript{st} Leonie Kallabis}
\IEEEauthorblockA{\textit{Institute of Computer Science} \\
\textit{University of Appl. Sc. Cologne}\\
Köln, Germany \\
leonie.kallabis@smail.th-koeln.de}
\and
\IEEEauthorblockN{2\textsuperscript{nd} Timo Bertram}
\IEEEauthorblockA{\textit{Institute for Machine Learning} \\
\textit{Johannes Kepler University}\\
Linz, Austria \\
bertram@ml.jku.at}
\and
\IEEEauthorblockN{3\textsuperscript{rd} Florian Rupp \thanks{*All authors contributed equally to this work.}}
\IEEEauthorblockA{\textit{Department of Computer Science} \\
\textit{Tech. University of Appl. Sc. Mannheim}\\
Mannheim, Germany \\
f.rupp@hs-mannheim.de}
}

\maketitle

\begin{abstract}


The visual style of game elements considerably contributes to the overall experience. Aesthetics influence player appeal, while the abilities of game pieces define their in-game functionality.
In this paper, we investigate how the visual style of collectible cards influences the players' perception of the card's actual strength in the game.
Using the popular trading card game Magic: The Gathering, we conduct a single-blind survey study that examines how players perceive the strength of AI-generated cards that are shown in two contrasting visual styles: cute and harmless, or heroic and mighty. 
Our analysis reveals that some participants are influenced by a card's visual appearance when judging its in-game strength.
Overall, differences in style perception are normally distributed around a neutral center, but individual participants vary in \emph{both} directions: some generally perceive the cute style to be stronger, whereas others believe that the heroic style is better.

\end{abstract}

\begin{IEEEkeywords}
Game perception, cognitive bias, visual style, collectible card games, games user research 
\end{IEEEkeywords}

\section{Introduction}
The visual style of a game is defined as ``a cohesive and unifying visual aesthetic"~\cite {donovan2013pretty} that concentrates on the visual perception of a pleasing appearance or beauty~\cite{Tractinsky2014}. Apart from the visual look, the style of games and game elements frequently provide insight into the functionality and expected usage~\cite{liapis2014computational}.

Determining how the visual style affects players comes with its own set of challenges. Individuals come from diverse backgrounds and bring unique preferences and experiences, making it challenging to accurately predict their perceptions. Consequently, player studies are a critical tool in human-computer interaction and game design to investigate player experience~\cite{drachen2018games,ashby_personalized_2023,rogers_using_2023,rupp2024balance}.
Despite the widely recognized importance of providing a pleasing visual experience to players, little research has been done on the impact of explicitly chosen styles on player perception. When players expect the look of game elements to provide hints about their usage, biases can shape how players interact with the game pieces. 

\begin{figure}[h]
  \centering
  \begin{subfigure}[b]{0.45\columnwidth}
    \centering
    \includegraphics[width=\textwidth]{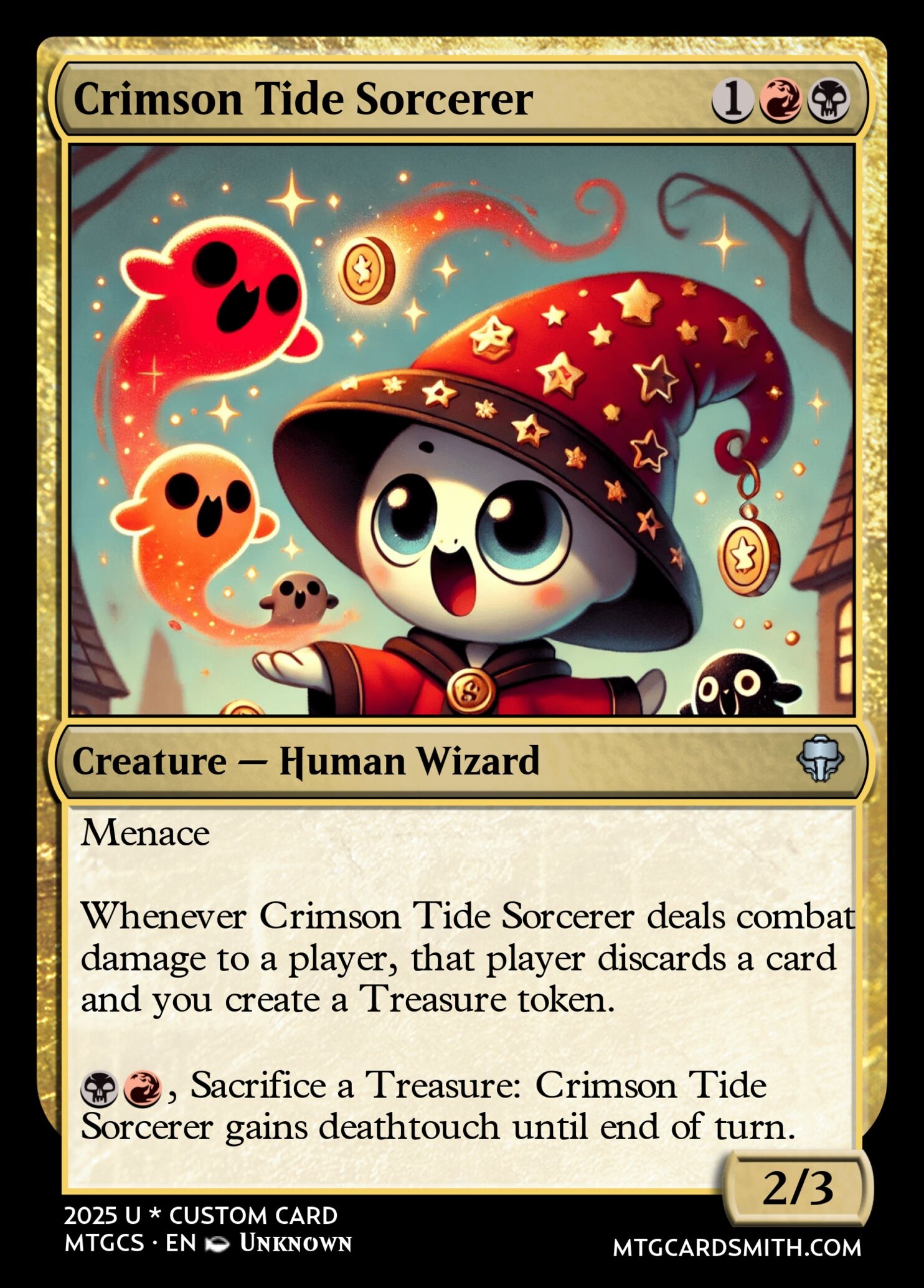} 
    \caption{A card in a \textit{cute} style.}
  \end{subfigure}
  \hfill
  \begin{subfigure}[b]{0.45\columnwidth}
    \centering
    \includegraphics[width=\textwidth]{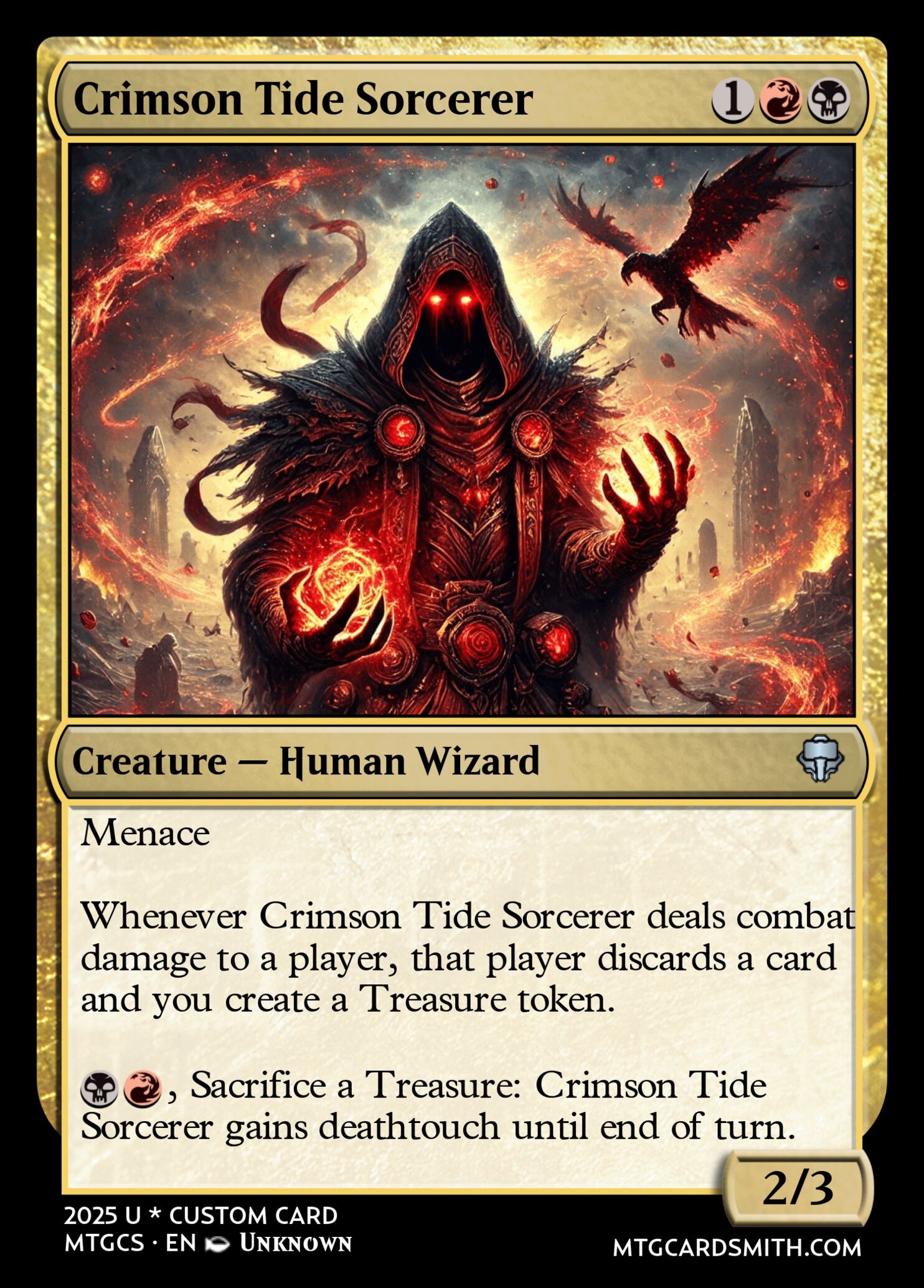} 
    \caption{A card in a \textit{heroic} style.}
  \end{subfigure}
  \caption{Comparing two different versions of \textit{card 11} in two different visual styles. The card's abilities and attributes remain identical, only the artwork varies.}
  \label{fig:card11}
  \vspace{-4mm}
\end{figure}

In this work, we investigate how the choice of visual style affects players' evaluations of the in-game strength of game elements. We focus on collectible cards from the game \textit{Magic: The Gathering (MTG)} as our game elements, using different styles for the same cards (cf. Figure~\ref{fig:card11}). Collectible card games like \textit{Magic: The Gathering} provide an excellent environment to test the influence of style on player perception because they naturally feature a diverse range of styles and task players to accurately evaluate cards. To test our hypothesis of style influencing strength perception, we conduct a survey where participants are shown a card in one of three styles - \textit{cute}, \textit{heroic}, or \textit{no style} - and are tasked to rate the strength of that card in the game. We speculate that players might be impacted by the style of a card, e.g., implicitly being biased to believe that \textit{cute} cards are weaker than \textit{heroic} ones.

We create new, unseen artworks of cards using recent advances in generative artificial intelligence (GenAI). With a text-to-image model, we generate artworks in different styles for the same card, using the card's text, abilities, and style-specific terms as a prompt.

\noindent In summary, our contributions are:
\begin{itemize}
    \item The design and conduction of a single-blind study to investigate card perception depending on visual style.
    \item A contribution to the understanding of visual style perception and GenAI utility in the context of MTG with:
    \begin{itemize}
        \item A comprehensive descriptive and statistical analysis of the survey data.
        \item A qualitative analysis of the generated cards.
    \end{itemize}
\end{itemize}

We give a detailed description of the game MTG in Section~\ref{sec:MTG}, before covering the related work (Section~\ref{sec:related}) and explaining the methodology (Section~\ref{sec:methods}). Finally, we summarize our results in Section~\ref{sec:results}, followed by limitations and conclusions. All data from this study, such as the generated cards and the survey responses, are publicly available.\footnote{\url{https://github.com/FlorianRupp/mtg-visual-card-style-survey}}


\section{Magic: The Gathering}
\label{sec:MTG}

\begin{figure}[h]
    \centering
    \includegraphics[width=\columnwidth]{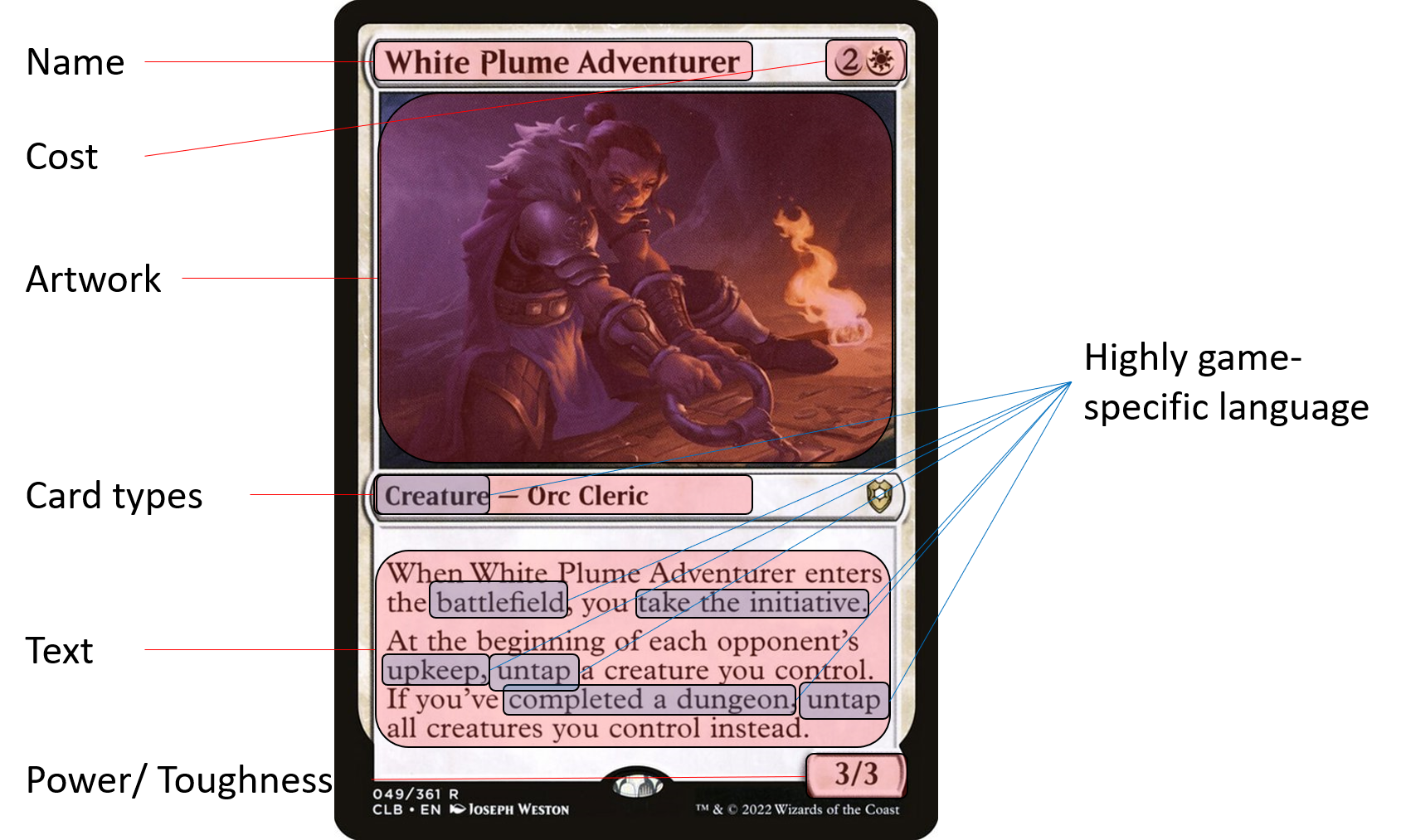}
    \caption{An exemplary card from MTG: Cards have different attributes that describe the functionality of a card, including numerical values, categories, and natural language text. The card text in particular is central to a card's function but contains game-specific terms that require a deep understanding of the game's rules.}
    \label{fig:mtg_card}
\end{figure}

\textit{Magic: The Gathering} is a prominent collectible card game (CCG), invented in 1993. Although MTG inspired a whole genre of card games, it has remained relevant to this day. One of MTG's main appeals is the constant release of novel cards that are developed for each new \textit{expansion}, or \textit{set}, of the game. Such \textit{sets} are released multiple times a year with 200-300 cards, the majority of which are completely novel. In the 32 years since its invention, MTG has released approximately 30,000 unique cards.\footnote{https://scryfall.com/search?q=-is\%3Areprint} Every card has its own name, artwork, attributes, and rule text (see Figure~\ref{fig:mtg_card}). The majority of this rule text uses game-specific language and terms. As such, MTG already has a large game complexity based on the cards that it is played with, notwithstanding the actual rules of gameplay.

MTG is a sequential, turn-taking, multiplayer game. Before playing the game, players decide on a subset of cards they want to play with their \textit{decks}. Throughout one game, players draw cards from their respective decks, creating a highly asymmetric game, in which players play against each other with different cards. Cards differ in strength, i.e., some cards can be strictly better than others. \textit{Strength} refers to the utility of a card in the game and how much impact it has on the gameplay. Stronger cards can cost less, have more or better abilities, be more powerful, or synergize better with the deck of a player. 

Due to the difference in card strength, only a small subset of cards in the game is used in competitive play. Still, competitive play rarely converges to all players using the same deck due to personal preferences, different priorities, and unique assumptions. In addition, some decks can inhabit \textit{Rock-Paper-Scissors}-esque behavior, where a deck is specifically built to be good against a different deck. This leads to metagaming decisions that result in an equilibrium of card choices.

Importantly, competitive play is structured into different formats that allow a different subset of cards to be played with. Some formats focus on newer cards, while others allow the majority of the card pool to be used. The strength of a card is highly dependent on the format it is evaluated in. In this work, we regard the strength of evaluated cards in comparison to real cards that have been released in the last months, therefore focusing less on formats that use older cards.

With the game outlined, we now summarize related work in the areas of CCGs and visual style. 

\section{Related Work}
\label{sec:related}

\noindent\textbf{Collectible Card Games:} Research on CCGs mostly focuses on creating sophisticated gameplay agents rather than the human perception of the game elements. Proposed competitions \cite{kowalski2023SummarizingStrategyCard, kowalski2024IntroducingTalesTributea, dockhorn2019IntroducingHearthstoneAICompetitiona} provide incentives to improve the current state of CCG agents for various games. This leads to different approaches of playing agents, such as symbolic agents \cite{stiegler2018SymbolicReasoningHearthstonea}, search-based agents \cite{choe2019EnhancingMonteCarloa}, and reinforcement learning agents \cite{xiao2023MasteringStrategyCard}. Work in the area of deck building focuses on the card choices of players rather than gameplay and utilizes contrastive learning \cite{bertram2021PredictingHumanCard, bertram2024LearningGeneralisedCard}, reinforcement learning \cite{esilvavieiraExploringReinforcementLearning2023}, or evolutionary algorithms \cite{kowalski2020EvolutionaryApproachCollectible}.


\noindent\textbf{Visual Style and Player Perception:} A game comprises various elements that are intricately designed, ultimately influencing its visual appearance. Although these elements, such as characters and objects, are similar across different games, their style can vary considerably. Over the years, research has attempted to define this concept and build a vocabulary to describe the visual style of games. Broadly, three primary styles have emerged: abstract, stylized, and realistic~\cite{jarvinen2002gran, McLaughlin2010}, which have been further explored and expanded upon in subsequent studies~\cite{donovan2013pretty,cho2018realism}. Game visual research compares the effects of different visual styles~\cite{gerling2013effects}, investigates the influence of individual visual elements like color, shape, or dimensionality~\cite{kallabis2024colorEmotion, garver2018impact, plass2020colorShapeDimensionality}, examines player perspectives~\cite{denisova2015first, garver2018impact}, analyzes character design~\cite{pradantyo2021antagonists, poeller2024}, and focuses on the influence of visual embellishments~\cite{hicks2019juicy, meiners2025}.

Visual embellishments have been shown to enhance the visual appeal and immersive experience of games~\cite{hicks2019juicy}. Lushness, which involves visual elements independent of player actions, positively affects player experience and visual appeal but does not significantly improve immersion~\cite{meiners2025}. 
Color has been a prominent aspect of investigation with heterogeneous results. Some suggest that color has minimal impact on the player experience~\cite{garver2018impact}, particularly concerning player emotions~\cite{kallabis2024colorEmotion}, others find color effective in inducing positive or negative emotions~\cite{plass2020colorShapeDimensionality} or enhancing immersion~\cite{Roohi2019}. Dimensionality and shape have been effectively utilized as emotion elicitors~\cite{plass2020colorShapeDimensionality}.
A player's perspective in a game (1st person, 3rd person) can affect a player's immersion~\cite{denisova2015first} and has limited impact on the player experience~\cite{garver2018impact}. 
Character perception is another way to influence the game experience, as the design of non-player characters can sway perception, such as them being seen as evil~\cite{pradantyo2021antagonists}, while the player character's looks can reflect their desired experience. A ``cute" and approachable appearance is valued more by players with affiliation motives, whereas those motivated by power prefer a strong, striking appearance~\cite{poeller2024}.
Surprisingly, despite the longstanding curiosity in understanding visual style in games~\cite{jarvinen2002gran}, there is limited research focusing on visual style comprehensively. When comparing abstract and stylized graphics, Gerling et al. found that stylized graphics generally result in a more favorable impression of the game, although this effect is dependent on the complexity of the game's mechanics~\cite{gerling2013effects}. This suggests that simpler games can still provide a positive experience despite abstract graphics.

\noindent\textbf{Our Work:} We aim to extend the current research by adding the perspective of how strength perception can be influenced by visual style. In the following section, we explain the methodology of our study.  


\section{Methodology}
\label{sec:methods}

Our methodology consists of three key steps: (1) selecting visual styles, (2) generating cards using text and image models, and (3) designing the study.

\subsection{Selection of Visual Styles}

Our goal is to identify defining properties for generating images that are both feasible in games and align with MTG's visual style. Although the perception of images is inherently subjective to some degree~\cite{cho2018realism}, we aimed to provide distinct descriptions based on previous research and implemented a manipulation check to ensure alignment with participants' perceptions.
Following the definition of visual style as cohesive~\cite{donovan2013pretty} and the three general types of visual style, abstract, stylized, and realistic~\cite{jarvinen2002gran, McLaughlin2010}, we select two distinct and contrastive styles to ensure clear recognition: one style aims for a powerful perceptual impact, while the other aims to be perceived as harmless.

Many MTG cards represent some kind of character. 
Depending on a character's posture and body shape, a character can be perceived as stronger or weaker~\cite{wellerdiek2015strong}. In addition, the proportions of a character play an important role. As Medley et al. summarize, body proportions affect perceived cuteness~\cite{cheok2012kawaii}, while unnaturally large heads or eyes can make a character appear cuter~\cite{medley2020cute}. 

Moreover, color can play a crucial role in perception, such as the affect~\cite{elliot2019color}. While strict mappings of color preferences require careful consideration, certain tendencies can still be observed~\cite{elliot2019color}.
Cheok and Fernando found that participants tend to associate pure hues, bright colors, and primary colors with cuteness, while darker shades could be perceived as mysterious~\cite{cheok2012kawaii}.
Since the harmless style uses characters with unnatural proportions, such as large heads, we chose the second style to represent realism in the context of games. We base this on a definition of realism as ``a style portraying characters and environments by attempting to achieve visual parity with real-world references.''~\cite{cho2018realism} This allows the artwork to depict fantasy creatures that could exist in the real world.

We derive properties for two contrasting art styles, which we refer to as \textit{cute} and \textit{heroic}:
\begin{itemize}
    \item \textbf{Cute:} Big eyes, large head, round face, happy expression, rounded and soft shapes~\cite{medley2020cute}; bright and pure or saturated colors~\cite{cheok2012kawaii}; stylized~\cite{cho2018realism, McLaughlin2010}
    \item \textbf{Heroic:} saturated colors~\cite{cheok2012kawaii}; strong body~\cite{wellerdiek2015strong}; realism~\cite{cho2018realism, jarvinen2002gran, McLaughlin2010}
\end{itemize}
Figure~\ref{fig:card11} shows an example of both the \textit{heroic} and \textit{cute} visual style displayed on a card used in the survey. The character in the cute style has a prominent big head, big eyes, and a happy expression. The shapes are mostly soft and the colors are mostly bright and saturated. The heroic style features a strong-bodied, realistic character presented in saturated colors.

\subsection{Card Generation}

We generate new MTG cards using generative AI models in a twofold process. First, we generate a card's description, including its name, ability, and overall stats. Secondly, we use the generated text to prompt a text-to-image model to create visually distinct card images in different styles.

\paragraph{Text Generation} Multiple large language models (LLMs) are used to design the card descriptions. We use the prompt: \textit{You are a collectible card game designer and are tasked to create new Magic: The Gathering cards. You are free to design the cards but they should adhere to conventions and rules in Magic: The Gathering. Design 3 different cards.}. To reduce bias related to any specific language model, we use 5 different models and pool the results. The chosen models are GPT4o \cite{achiam2023gpt}, Claude 3.5 Haiku \cite{anthropic2024claude}, GPTo1\footnote{At the time of writing, no official paper has been published on the o1 models.}, GPTo1-mini, and DeepSeek-R1 \cite{guo2025deepseek}.

\paragraph{Image Generation}

In the next step, we generate artwork for the previously generated cards using DALL-E 3 \cite{betker2023improving} with prompts specifying the particular style and the card's text. We use the prompt: \textit{I need an artwork for Magic: The Gathering. You should only draw the card image, not the text of the card! It is important that the card is in a [cute, friendly, harmless OR mighty, heroic, powerful] style. This is the card: [CARD TEXT]}. Providing the full card text allows the models to adjust the image based on the card mechanics, which might result in outputs more closely matching the intention behind the card design.

\paragraph{Final Card Generation}

We use a tool to create the final MTG cards shown to participants\footnote{\url{mtgcardsmith.com}}. The LLMs provide all card features, so no changes or additions are required to create plausible cards. 

\subsection{Study Design}

A key contribution of this paper is the design of a study that is structured into five parts: definition of measures, data collection, participant description, manipulation check, and overall procedure. Figure~\ref{fig:procedure} provides an overview.

\subsubsection{Measures}
We formulate the study design according to the research question \textit{How does the choice of visual style impact the perceived strength and card popularity?} We designate the visual style as the independent variable and account for the card variant (excluding the visual style) to ensure the card content remains independent of the visual style. One dependent variable, the perceived strength is set.

This study employs a single-blind, repeated-measures within-subject design. To compensate for the effects of possible subject bias, study participants were divided into three groups. Each group was presented with different combinations of card variants and visual styles. We chose this methodological approach to minimize the chance that participants would infer the experimental hypothesis and to minimize bias in their responses.

\subsubsection{Data collection}
All data is self-reported and each dependent variable is assessed using a single questionnaire item.
To measure the perceived strength of cards, participants are asked to compare the card to real cards from recent standard-legal sets. Responses are recorded on a 7-point Likert scale.

\subsubsection{Participants}
A total of 65 participants completed the survey, although three were excluded from further analysis as their completion time was under six minutes, which we classified as speeders. The survey was distributed online, reaching tournament players, MTG player communities, and a pool of students.

The participants' ages range from 20 to 52 years, with an average age of 31 and a standard deviation of 6.21. The gender distribution within the participant group shows a predominance of males, with 55 identifying as male, 4 as female, 2 as non-binary, and 1 participant not specifying their gender.
Most participants have a high level of expertise, with 41.9\% reporting that they play the game regularly and 43.5\% identifying themselves as competitive tournament players. Additionally, a majority of participants have been playing MTG for an extended period, with 77\% of the individuals reporting more than 5 years of experience. To ensure participants could comprehend the information displayed on the cards and accurately assess their perceived strength, proficiency in playing MTG was required. Participants were only allowed to continue the study if they confirmed their proficiency in the game. 



\begin{figure}
    \centering
    \includegraphics[width=1\linewidth]{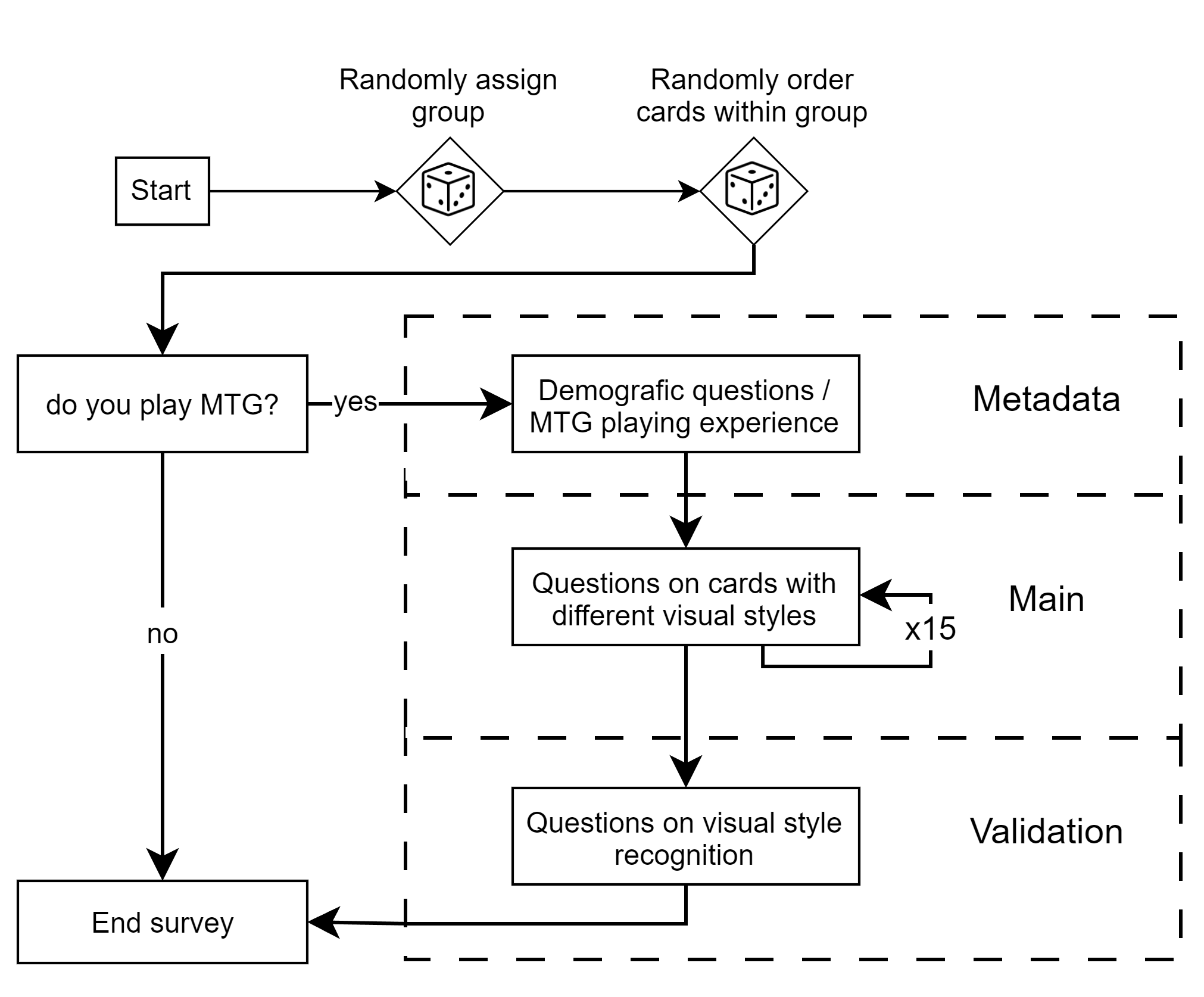}
    \caption{The survey randomly assigns participants to one of three groups, which are shown cards in different styles. The survey consists of three sections: metadata collection, main evaluation, and validation questions.}
    \label{fig:procedure}
\end{figure}

\subsubsection{Manipulation Check}
We implement a manipulation check (see Figure~\ref{fig:procedure}) at the end of the survey to ensure that participants can distinguish between the visual styles presented. Participants are shown four consecutive, randomized images — two with a cute visual style and two with a heroic visual style. 
For each image, participants are asked to select any number of words from a list of six (cute, friendly, harmless, mighty, heroic, and powerful) that best describe the image in their opinion. These words were drawn from the prompts used for the visual styles.

We score participants' responses as follows: one point for each correctly identified word and one point for not selecting any incorrect words. Thus, participants could earn a maximum of eight points across the four images. The majority (93.6\%) performed well, with 48.4\% scoring 8 points, 30.7\% scoring 7 points, and 14.5\% scoring 6 points. This indicates that the participants were able to differentiate between the two visual styles. 


\subsubsection{Procedure}
The overall procedure of the survey is illustrated in Figure~\ref{fig:procedure}.
Upon initiating the online survey, participants are provided with information about the survey's purpose, which is distinct from the actual hypothesis to maintain blindness and prevent bias. Participants were told that the survey was about how well GenAI can create MTG cards.
Participants are first asked whether they play MTG; those who respond affirmatively proceed to the main part of the study, which includes demographic questions and inquiries about their MTG experience (Step: Metadata).
Thereafter, the primary data collection process begins, consisting of 15 consecutive pages. Each page displays a randomly selected card, questions about the dependent variables, and a free-text option for additional feedback. 
After completing the 15 pages, participants proceed to the manipulation check, where they assess the visual style recognition (Step: Validation).

\section{Results and Discussion}
\label{sec:results}

For analysis, we only include fully completed submissions, resulting in 62 total samples. With 15 cards evaluated in three different styles (cute, heroic, no image), this amounts to 310 ratings per style.

In the following, we analyze the collected data to explore how perceived card strength varies with visual style across different dependent variables, such as cards and players.
To address these questions, we will combine descriptive insights with hypothesis testing.


\subsection{Style in general: Do players generally misjudge dependent on visual style overall?}
\label{sec:results-general}

We answer this question by comparing the data distributions of participants' ratings of the perceived strength of all cards dependent on the visual style. After a descriptive comparison, we evaluate the difference of the distributions with hypothesis testing.

\begin{figure}
    \centering
    \includegraphics[width=01\linewidth]{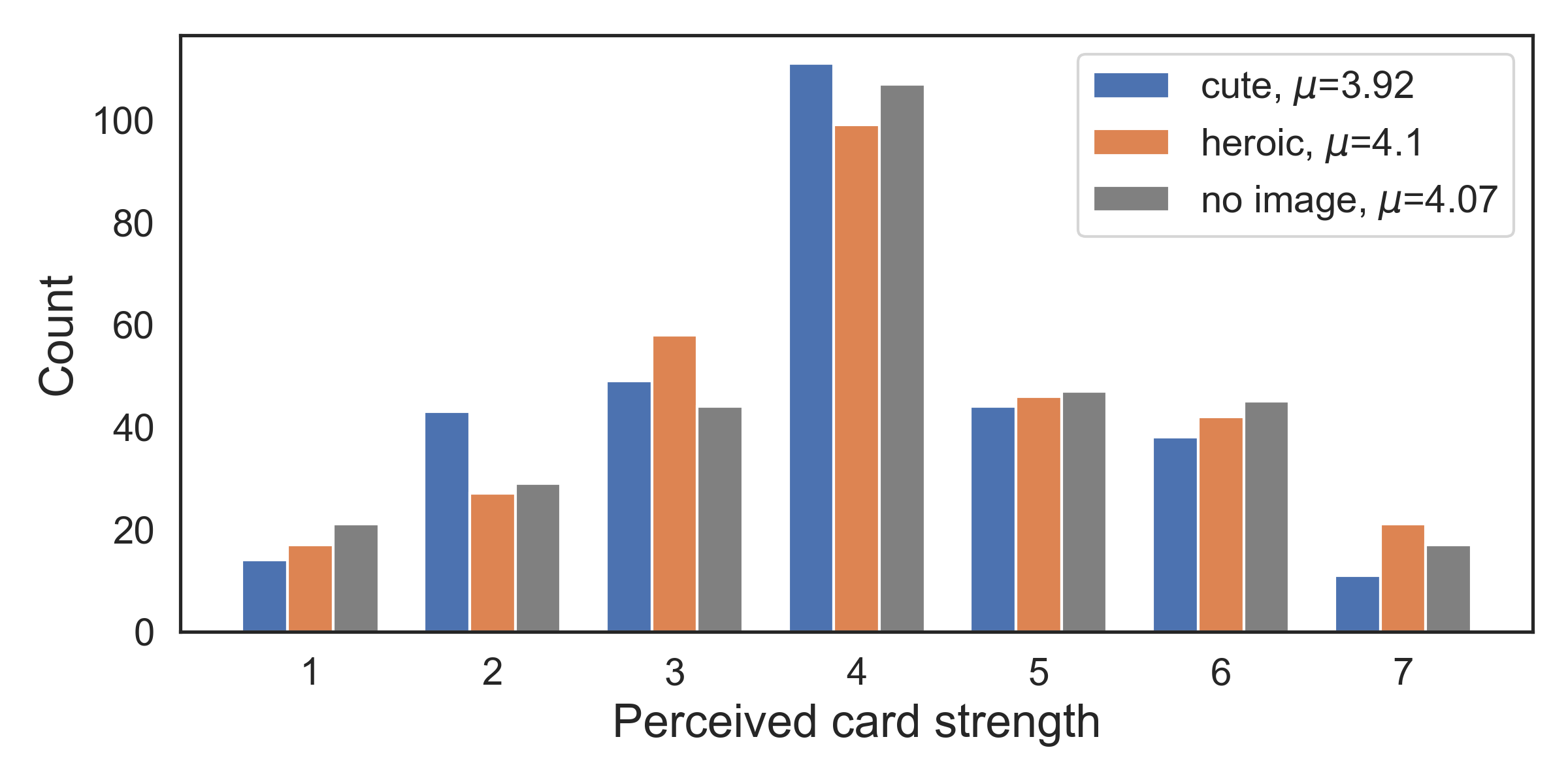}
  \caption{The distribution of perceived strength per visual style, independent of a particular card or participant. The rating 1 refers to \emph{too weak}, 4 to \emph{balanced}, and 7 to \emph{too strong}.}
    \label{fig:card-ratings-overall}
\end{figure}

A histogram of the three distributions of card ratings on strength per visual style is presented in Figure~\ref{fig:card-ratings-overall}. We observe a marginally higher mean rating for cards in the heroic style (4.1) and cards with no image (4.07) than in the cute style (3.92). Nearly twice as often, a card with a heroic style received the highest strength rating compared to a card with a cute image style (21 times and 11 times). However, comparing the means of the distribution, we find no effect using Cohen's~D (0.12) \cite{cohen_statistical_2013}. 


To further investigate disparities in the distributions, we define the following hypotheses:
\begin{itemize}
    \item $H_0$: There is no overall significant difference in perceived card strength based on the style.
    \item $H_1$: There is an overall significant difference in perceived card strength based on the style.
\end{itemize}

We use the Mann-Whitney-U test for hypothesis testing of unpaired, ordinal data. With a p-value of 0.18, we accept $H_0$, stating that the data distributions are statistically \emph{not} different. Therefore, we conclude that no visual style deceived participants to rate card strength differently.

In addition, we asked participants if they would like to own each card as a yes-or-no question to determine preference. Interest in the cards was generally low, with no significant trend based on preferred style (37\% cute, 35\% no image, and 35\% heroic).

\subsection{Cards and Style: Are there particular cases where players misjudge?}

Since we previously concluded that there was no significant difference in participants' perceptions based solely on visual style, we further explore this by adding cards as a dependent variable. We follow the same analysis steps as in the previous section. 

Table~\ref{tab:cards-cohensd} compares the perceived ratings of all visual style combinations per card, counting for how many cards a particular effect is observed.
In contrast to the results in Section~\ref{sec:results-general}, we find a small effect for approximately half of all cards for all visual style comparisons. A medium effect can be observed for a minority of two cards when comparing cute and heroic cards with no visual style. No large effect exists for any card.

\begin{table}[]
\centering
\caption{Comparing visual style perception on card strength dependent on single cards. For each style comparison, we report the number where an effect according to Cohen's D can be observed.}
\begin{tabular}{lccc}
\toprule
 & \multicolumn{3}{c}{Number of cards with Cohen's D effect:}  \\
 \textbf{visual style} & \textbf{small} ($\geq 0.2$) & \textbf{medium} ($\geq 0.5$) & \textbf{large} ($\geq 0.8$) \\
 \cmidrule(lr){1-1} \cmidrule(lr){2-2}  \cmidrule(lr){3-3}  \cmidrule(lr){4-4}
 cute vs. heroic     & 8 & 0 & 0 \\
 no image vs. heroic & 7 & 2 & 0 \\
 cute vs. no image   & 8 & 2 & 0 \\ \bottomrule
\end{tabular}
\label{tab:cards-cohensd}
\end{table}

To test if any distribution of perceived card strength varies, we apply the same hypothesis tests as in the previous section. Therefore, we define the following hypothesis for all 15 cards:
\begin{itemize}
    \item $H_0$: For a selected card there is no significant difference in perceived card strength based on the style.
    \item $H_1$: For a selected card there is a significant difference in perceived card strength based on the style.
\end{itemize}

Except for card 11, all p-values are greater than 0.05 and thus we accept $H_0$. For card 11 (Figure~\ref{fig:card11}), however, cards with no image were perceived significantly different than cute (p=0.036) and heroic (p=0.027) visual style cards. With an average strength rating of 5.1, this card with no image was perceived stronger by participants than the same card with a cute (4.32) or a heroic (4.37) image.

Based on Cohen's D values and hypothesis testing, we conclude that while style generally does not affect overall player perception, it can influence how players perceive the balance of particular cards. While the effect remains small for about half of the tested cards, one card shows a significant difference when no image is present.

\subsection{Participants and Style: How does card strength rating depend on participants?}
\label{sec:results:participants-style}

In this section, we go a step further and examine how participants perceived a card's strength based on its visual style, independent of any specific card. Specifically, we analyze how much a participant's average deviation from the overall rated strength differs depending on the visual style. Since some cards are inherently rated as stronger or weaker, we apply a normalization of ratings to isolate a participant's deviation from the card's actual strength.
The average deviation of a participant's rating $p$ of the set of all cards $C$ of the same style $s$ the participant has seen is computed with Equation~\ref{eq:user-style-perception}. $r_{c,p,s}$ is the rating of a selected participant for a selected card $c$ for a given style; $\bar{r_c}$ is the overall average rating of a selected card.

\vspace{-2mm}
\begin{equation}
\label{eq:user-style-perception}
    rating(p, s) = \frac{1}{|C|} \sum^{C}_{c\,\in\,C} r_{c,p,s} - \bar{r_c}
\end{equation}


\begin{figure}
    \centering
    \includegraphics[width=\linewidth]{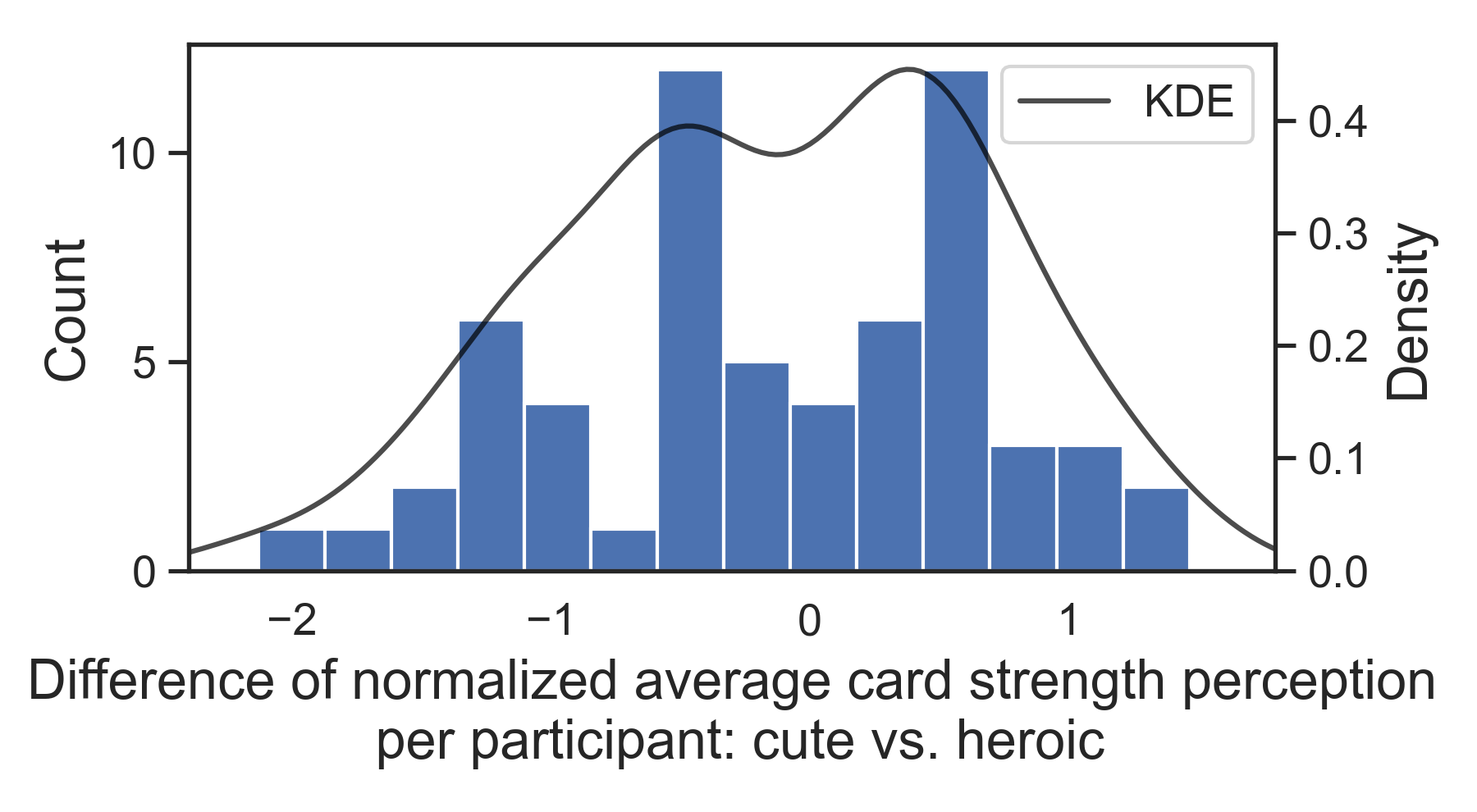}
  \caption{How do participants differ in their perceptions of strength in the cute and heroic visual style? The distribution of the difference of the normalized average rating per participant using Eq.~\ref{eq:user-style-perception} is displayed. The kernel density estimation hints at two smaller peaks within the distribution: one for participants tending to prefer the cute visual style ($<0$) and vice-versa ($>0$).}
    \label{fig:card_rating_per_user}
    \vspace{-5mm}
\end{figure}


To investigate the overall difference in perceptions of the cute and heroic visual style per participant, we compute $rating(p, s=heroic) - rating(p, s=cute)$ for all participants.
The resulting data is presented in the histogram along with a kernel density (KDE) plot in Figure~\ref{fig:card_rating_per_user}. The more a participant is positioned away from the equilibrium (0), the more the cute style ($<0$) is favored over the heroic style and vice versa ($>0$).
We further observe two small density peaks in the KDE plot, which is backed up by the histogram. This hints that the data might be bimodal, with one group of participants being deceived to perceive cards in a cute visual style as stronger than heroic ones and vice versa.

A Shapiro-Wilk test reveals that the data is normally distributed (p=0.44) and a Gaussian mixture model proves that a single normal distribution can explain the data better than two.
While we cannot statistically assume that there are two separated groups, we see that participants' perceived difference between the two visual styles goes in \emph{both} directions with small density peaks for both sides, not only in the direction of \emph{one} single visual style.

That being said, this further explains our finding in Section~\ref{sec:results-general}, where we cannot see larger differences in the mean values (cf. Figure~\ref{fig:card-ratings-overall}). Since we observe that participants' deception for a style is normally distributed around a neutral center, no particular visual style has a dominant effect.

The small density peaks could hint that different preferences in visual style affect strength perception. Moreover, the number of participants who prefer a distinct visual style is quite similar for cute and heroic.
Building on Bartle's notion of clustering players based on their in-game behavior~\cite {bartle1996hearts}, visual style might further distinguish player categories. Some players may prioritize visual appeal more than others, segmenting them based on their stylistic preferences. This indicates a possible avenue for further exploration of how visual aesthetics may impact player engagement and behavior.

\section{Further Analysis}

With this main focus outlined, we provide further analyses of our obtained data. This section focuses on three auxiliary questions:

\begin{itemize}
    \item What problems can we find in the generated cards?
    \item How consistent are players at rating the strength of cards?
    \item What are the participant's general responses to the generated cards?
\end{itemize}

\subsection{Errors in Generated Cards}

Cards in MTG are carefully designed to fit the game's balancing, thematics, and narrative. A large number of game elements have clearly defined characteristics that players are used to, and which need to be consistent across new card releases. Interestingly, the LLMs used for card generation do not fully capture these underlying world rules and create cards that break them.

\begin{figure}[htbp]
  \centering
  \begin{subfigure}[b]{0.45\columnwidth}
    \centering
    \includegraphics[width=\textwidth]{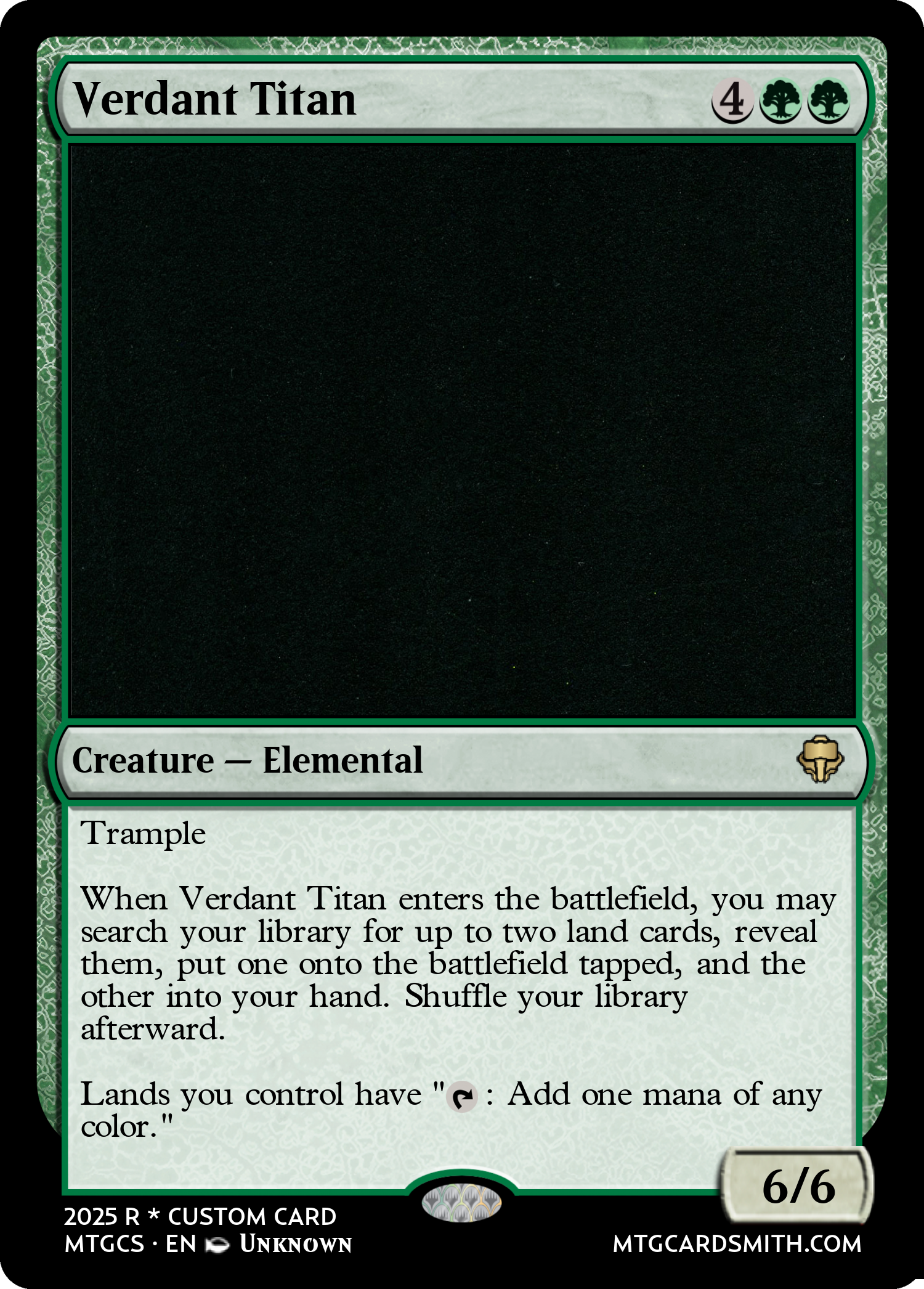} 
    \caption{LLM-generated card}
  \end{subfigure}
  \hfill
  \begin{subfigure}[b]{0.45\columnwidth}
    \centering
    \includegraphics[width=\textwidth]{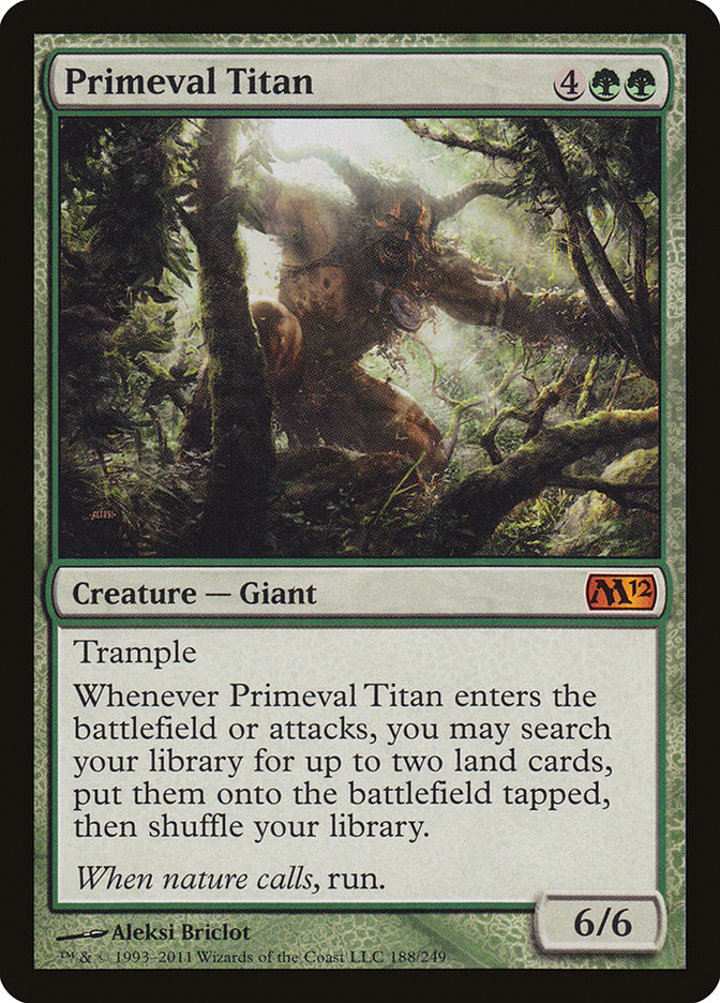} 
    \caption{Real MTG card}
  \end{subfigure}
  \caption{Some LLM-generated cards closely resemble existing MTG cards, which can make them difficult to use for game designers. Additionally, it strongly hints that the LLM was trained on MTG cards.}
  \label{fig:primeval_titan_comparison}
\end{figure}

We found several types of issues with LLM-generated cards:

\begin{itemize}
    \item Some cards are very similar to existing cards. For example, the card \textit{Verdant Titan} is a very close resemblance to the real card \textit{Primeval Titan} (Figure~\ref{fig:primeval_titan_comparison}) in color, power, and text, but even in name.
    \item Some game elements' classical functionality is changed. For example, a wolf token (Card 8)\footnote{See related GitHub repository: https://github.com/FlorianRupp/mtg-visual-card-style-survey.\label{sharednote}} should not gain the player life points when dying.
    \item The functions of some cards are not well-aligned with the game's narrative. For example, wizards (Card 11)\footnoteref{sharednote} rarely have an effect when they deal damage to the opponent. 
    \item The different abilities on a card do not work together as players will likely expect. Card 14\footnoteref{sharednote} creates a creature token\footnote{Creature tokens are not cards in a player's deck, but supplementary game elements that can be created by cards} every turn and has the additional ability that the player is allowed to return creatures to their hand when they die. However, the game's rules say that creature tokens can not be in a player's hand. The card highly suggests that the two abilities synergize but the rules prevent this. We see this break of expectations in our survey, as one participant seemed to not be aware of the rule interaction and specifically mentioned that this would be too strong.
\end{itemize}

While all of these can easily be fixed by game designers, it requires careful investigation. An experienced designer will intuitively understand what card features match the game's history and atmosphere, but such errors can easily be missed.

\subsection{Heterogeniety of Card Ratings}

\begin{figure}[h]
    \centering
    \includegraphics[width=\columnwidth]{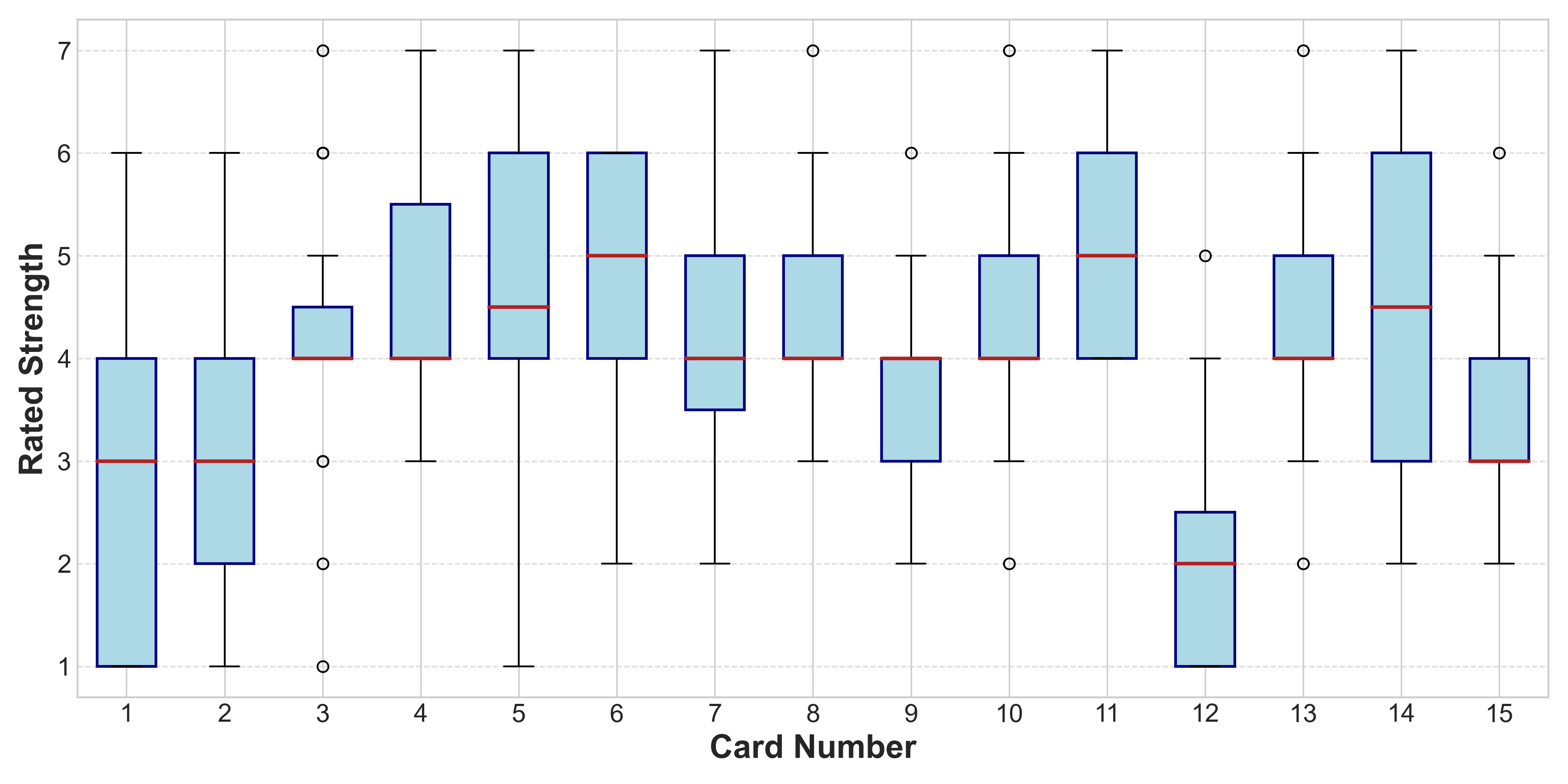}
    \caption{Per-card distribution of participant's strength rating for cards with no image. In some cases, we find a large variance in how self-reported expert players rate the strength of the same card, showing the difficulty of the task.}
    \label{fig:rating_distribution}
\end{figure}

Lastly, we find that players have vastly different perceptions of the individual strength of cards. Although the majority of players reported to be experts, the inter-quartile range (Figure~\ref{fig:rating_distribution}) frequently spans 2 to 3 points, i.e., the difference between a rating of \textit{basically useless} and \textit{balanced} strength. We only include ratings for cards without images to avoid visual deception.
 
This underlines the difficulty of evaluating collectible cards, and how players' differing views lead to different perceptions of cards. In the real-world development process, collectible cards often undergo rigorous, iterative testing processes, using the cards in actual gameplay to get a better understanding of their strength. 

\subsection{Participants' Opinions on GenAI}

We received numerous responses providing unprompted opinions on the use of GenAI for games like MTG. Participants did not show any explicit positive reactions, but multiple participants voiced unhappiness with the potential use of GenAI. This observation should alert stark caution in game designers debating its use in their products.

\section{Limitations}

This study surveys MTG players, though recruitment bias may limit the generalizability of the findings and their transferability.
Although we distributed the survey across social media, playing groups, and universities, we received a rather homogeneous group of participants. The vast majority of participants were male (88\%), describing themselves as either regular (41.94\%) or competitive tournament (43.55\%) players, respectively.
While this sample can be considered representative of MTG players at this experience level, it remains unclear how our results apply to less experienced players, more gender-diverse groups, or other games and genres. 
There have been observations that competitive players often prioritize winning over having fun~\cite{paiva2018player}. This focus may lessen the effect the visual style can have on experienced players and further indicates the need to investigate effects on less experienced players.

In addition, we only regard collectible card ratings in a relaxed online survey situation. Participants might be affected differently in competitive real-world gameplay situations, which can be stressful and require more intuitive responses.


\section{Conclusion}
We conducted a single-blind study to examine how visual style affects perceived in-game strength, using the trading card game \emph{Magic: The Gathering} (MTG) as a survey study. Participants in different groups evaluated AI-generated cards that varied only in style: cute, heroic, or no image.
While no overall bias towards a particular style was found, the effect depended on the individual case. For about half of the test cards, Cohen’s D indicated a small effect, suggesting that style influenced perception in these cases.
At the player level, normalized differences in perception followed a normal distribution centered around neutrality. However, participants showed deviations in \emph{both} directions, with some favoring cute cards and others heroic ones.

So can the visual style in game design create misleading perceptions of in-game strength in MTG? Based on our findings, we conclude \emph{yes slightly}, however, this is dependent on additional variables such as a particular card or player.
While our study design can be applied to other games, this study's participants do not resemble a representative sample to transfer findings to other game genres. 
In future work, we aim to address this by conducting a broader investigation with a more representative sample of players, in order to obtain more transferable results to other games and contexts.

\section*{Acknowledgements}
Timo Bertram acknowledges funding from COST Action CA22145, supported by COST (European Cooperation in Science and Technology).
Florian Rupp acknowledges funding from the Volkswagen Foundation (Project: Consequences of Artificial Intelligence on Urban Societies, Grant 98555).

    

\bibliographystyle{IEEEtranS}
\bibliography{bib}

\begin{thebibliography}{10}
\providecommand{\url}[1]{#1}
\csname url@samestyle\endcsname
\providecommand{\newblock}{\relax}
\providecommand{\bibinfo}[2]{#2}
\providecommand{\BIBentrySTDinterwordspacing}{\spaceskip=0pt\relax}
\providecommand{\BIBentryALTinterwordstretchfactor}{4}
\providecommand{\BIBentryALTinterwordspacing}{\spaceskip=\fontdimen2\font plus
\BIBentryALTinterwordstretchfactor\fontdimen3\font minus \fontdimen4\font\relax}
\providecommand{\BIBforeignlanguage}[2]{{%
\expandafter\ifx\csname l@#1\endcsname\relax
\typeout{** WARNING: IEEEtranS.bst: No hyphenation pattern has been}%
\typeout{** loaded for the language `#1'. Using the pattern for}%
\typeout{** the default language instead.}%
\else
\language=\csname l@#1\endcsname
\fi
#2}}
\providecommand{\BIBdecl}{\relax}
\BIBdecl

\bibitem{achiam2023gpt}
J.~Achiam, S.~Adler, S.~Agarwal, L.~Ahmad, I.~Akkaya, F.~L. Aleman, D.~Almeida, J.~Altenschmidt, S.~Altman, S.~Anadkat \emph{et~al.}, ``Gpt-4 technical report,'' \emph{arXiv preprint arXiv:2303.08774}, 2023.

\bibitem{anthropic2024claude}
A.~Anthropic, ``The claude 3 model family: Opus, sonnet, haiku,'' \emph{Claude-3 Model Card}, vol.~1, 2024.

\bibitem{ashby_personalized_2023}
T.~Ashby, B.~K. Webb, G.~Knapp, J.~Searle, and N.~Fulda, ``Personalized {Quest} and {Dialogue} {Generation} in {Role}-{Playing} {Games}: {A} {Knowledge} {Graph}- and {Language} {Model}-based {Approach},'' in \emph{Proceedings of the 2023 {CHI} {Conference} on {Human} {Factors} in {Computing} {Systems}}, ser. {CHI} '23, 2023, pp. 1--20.

\bibitem{bartle1996hearts}
R.~Bartle, ``Hearts, clubs, diamonds, spades: Players who suit muds,'' \emph{Journal of MUD research}, vol.~1, no.~1, p.~19, 1996.

\bibitem{bertram2021PredictingHumanCard}
T.~Bertram, J.~F{\"u}rnkranz, and M.~M{\"u}ller, ``Predicting human card selection in magic: {{The}} gathering with contextual preference ranking,'' in \emph{2021 {{IEEE Conference}} on {{Games}} ({{CoG}})}.\hskip 1em plus 0.5em minus 0.4em\relax IEEE, 2021, pp. 1--8.

\bibitem{bertram2024LearningGeneralisedCard}
------, ``Learning {{With Generalised Card Representations}} for ``{{Magic}}: {{The Gathering}}'','' in \emph{2024 {{IEEE Conference}} on {{Games}} ({{CoG}})}.\hskip 1em plus 0.5em minus 0.4em\relax IEEE, 2024, pp. 1--8.

\bibitem{betker2023improving}
J.~Betker, G.~Goh, L.~Jing, T.~Brooks, J.~Wang, L.~Li, L.~Ouyang, J.~Zhuang, J.~Lee, Y.~Guo \emph{et~al.}, ``Improving image generation with better captions,'' \emph{Computer Science. https://cdn. openai. com/papers/dall-e-3. pdf}, vol.~2, no.~3, p.~8, 2023.

\bibitem{cheok2012kawaii}
A.~D. Cheok and O.~N.~N. Fernando, ``Kawaii/cute interactive media,'' \emph{Universal Access in the Information Society}, vol.~11, pp. 295--309, 8 2012.

\bibitem{cho2018realism}
H.~Cho, A.~Donovan, and J.~H. Lee, ``Art in an algorithm: A taxonomy for describing video game visual styles,'' \emph{Journal of the Association for Information Science and Technology}, vol.~69, pp. 633--646, 5 2018.

\bibitem{choe2019EnhancingMonteCarloa}
J.~S.~B. Choe and J.-K. Kim, ``Enhancing {{Monte Carlo Tree Search}} for {{Playing Hearthstone}},'' in \emph{2019 {{IEEE Conference}} on {{Games}} ({{CoG}})}, Aug. 2019, pp. 1--7.

\bibitem{cohen_statistical_2013}
J.~Cohen, \emph{Statistical {Power} {Analysis} for the {Behavioral} {Sciences}}, 2nd~ed.\hskip 1em plus 0.5em minus 0.4em\relax New York: Routledge, May 2013.

\bibitem{denisova2015first}
A.~Denisova and P.~Cairns, ``First person vs. third person perspective in digital games: do player preferences affect immersion?'' in \emph{Proceedings of the 33rd annual ACM conference on human factors in computing systems}, 2015, pp. 145--148.

\bibitem{dockhorn2019IntroducingHearthstoneAICompetitiona}
A.~Dockhorn and S.~Mostaghim, ``Introducing the {{Hearthstone-AI Competition}},'' May 2019.

\bibitem{donovan2013pretty}
A.~Donovan, H.~Cho, C.~Magnifico, and J.~H. Lee, ``Pretty as a pixel: issues and challenges in developing a controlled vocabulary for video game visual styles,'' in \emph{Proceedings of the 13th ACM/IEEE-CS joint conference on Digital libraries}.\hskip 1em plus 0.5em minus 0.4em\relax ACM, 7 2013, pp. 413--414.

\bibitem{drachen2018games}
A.~Drachen, P.~Mirza-Babaei, and L.~E. Nacke, \emph{Games user research}.\hskip 1em plus 0.5em minus 0.4em\relax Oxford University Press, 2018.

\bibitem{esilvavieiraExploringReinforcementLearning2023}
R.~E~Silva~Vieira, A.~Rocha~Tavares, and L.~Chaimowicz, ``Exploring reinforcement learning approaches for drafting in collectible card games,'' \emph{Entertainment Computing}, vol.~44, p. 100526, Jan. 2023.

\bibitem{elliot2019color}
A.~J. Elliot, ``A historically based review of empirical work on color and psychological functioning: Content, methods, and recommendations for future research,'' \emph{Review of General Psychology}, vol.~23, pp. 177--200, 6 2019.

\bibitem{garver2018impact}
S.~Garver, N.~Adamo-Villani, and H.~Dib, ``The impact of visual style on user experience in games,'' \emph{EAI Endorsed Transactions on Game-Based Learning}, vol.~4, p. 153535, 1 2018.

\bibitem{gerling2013effects}
K.~M. Gerling, M.~Birk, R.~L. Mandryk, and A.~Doucette, ``The effects of graphical fidelity on player experience,'' in \emph{Proceedings of international conference on Making Sense of Converging Media}, 2013, pp. 229--236.

\bibitem{guo2025deepseek}
D.~Guo, D.~Yang, H.~Zhang, J.~Song, R.~Zhang, R.~Xu, Q.~Zhu, S.~Ma, P.~Wang, X.~Bi \emph{et~al.}, ``Deepseek-r1: Incentivizing reasoning capability in llms via reinforcement learning,'' \emph{arXiv preprint arXiv:2501.12948}, 2025.

\bibitem{hicks2019juicy}
K.~Hicks, K.~Gerling, P.~Dickinson, and V.~V. Abeele, ``Juicy game design: Understanding the impact of visual embellishments on player experience,'' in \emph{Proceedings of the Annual Symposium on Computer-Human Interaction in Play}.\hskip 1em plus 0.5em minus 0.4em\relax ACM, 10 2019, pp. 185--197.

\bibitem{jarvinen2002gran}
A.~J{\"a}rvinen, ``Gran stylissimo: The audiovisual elements and styles in computer and video games,'' in \emph{Computer games and digital cultures conference proceedings}, 2002.

\bibitem{kallabis2024colorEmotion}
L.~Kallabis, B.~Baruque-Zanón, H.~Klocke, A.~M. Lara-Palma, and B.~Naujoks, ``Investigating the effect of color stimuli on player emotions in games,'' in \emph{2024 IEEE Conference on Games (CoG)}.\hskip 1em plus 0.5em minus 0.4em\relax IEEE, 8 2024, pp. 1--4.

\bibitem{kowalski2023SummarizingStrategyCard}
J.~Kowalski and R.~Miernik, ``Summarizing {{Strategy Card Game AI Competition}},'' in \emph{2023 {{IEEE Conference}} on {{Games}} ({{CoG}})}, Aug. 2023, pp. 1--8.

\bibitem{kowalski2024IntroducingTalesTributea}
J.~Kowalski, R.~Miernik, K.~Polak, D.~Budzki, and D.~Kowalik, ``Introducing {{Tales}} of {{Tribute AI Competition}},'' in \emph{2024 {{IEEE Conference}} on {{Games}} ({{CoG}})}, Aug. 2024, pp. 1--8.

\bibitem{kowalski2020EvolutionaryApproachCollectible}
J.~Kowalski and R.~Miernik, ``Evolutionary approach to collectible arena deckbuilding using active card game genes,'' in \emph{2020 {{IEEE Congress}} on {{Evolutionary Computation}} ({{CEC}})}.\hskip 1em plus 0.5em minus 0.4em\relax IEEE, 2020, pp. 1--8.

\bibitem{liapis2014computational}
A.~Liapis, G.~N. Yannakakis, and J.~Togelius, ``Computational game creativity.''\hskip 1em plus 0.5em minus 0.4em\relax International Conference on Computational Creativity, ICCC, 2014.

\bibitem{McLaughlin2010}
T.~McLaughlin, D.~Smith, and I.~A. Brown, ``A framework for evidence based visual style development for serious games,'' in \emph{Proceedings of the Fifth International Conference on the Foundations of Digital Games}, ser. FDG '10.\hskip 1em plus 0.5em minus 0.4em\relax New York, NY, USA: Association for Computing Machinery, 2010, p. 132–138.

\bibitem{medley2020cute}
S.~Medley, B.~Zaman, and P.~Haimes, \emph{The Role of Cuteness Aesthetics in Interaction}, 2020, pp. 125--138.

\bibitem{meiners2025}
A.-L. Meiners, D.~Reich, K.~Hicks, D.~Alexandrovsky, and K.~Gerling, ``Lushness in game design: The role of non-interactive visual embellishments in player experience,'' in \emph{Proceedings of the 20th International Conference on the Foundations of Digital Games}, 2025.

\bibitem{paiva2018player}
F.~G. R.~M. Paiva, A.~d.~O. da~Rocha~Franco, G.~M. Junior, and J.~G.~R. Maia, ``Analyzing player profiles in collectible card games,'' \emph{Anais do XVII Simp{\'o}sio Brasileiro de Jogos e Entretenimento Digital-SBGames}, 2018.

\bibitem{plass2020colorShapeDimensionality}
J.~L. Plass, B.~D. Homer, A.~MacNamara, T.~Ober, M.~C. Rose, S.~Pawar, C.~M. Hovey, and A.~Olsen, ``Emotional design for digital games for learning: The effect of expression, color, shape, and dimensionality on the affective quality of game characters,'' \emph{Learning and Instruction}, vol.~70, 12 2020.

\bibitem{poeller2024}
S.~Poeller, N.~Baumann, and R.~L. Mandryk, ``Intimidating or friendly? how players represent themselves with character appearances that reflect their social motivations,'' in \emph{Proceedings of the 19th International Conference on the Foundations of Digital Games}.\hskip 1em plus 0.5em minus 0.4em\relax Association for Computing Machinery, 2024.

\bibitem{pradantyo2021antagonists}
R.~Pradantyo, M.~V. Birk, and S.~Bateman, ``How the visual design of video game antagonists affects perception of morality,'' \emph{Frontiers in Computer Science}, vol.~3, 4 2021.

\bibitem{rogers_using_2023}
K.~Rogers, V.~Le~Claire, J.~Frommel, R.~Mandryk, and L.~E. Nacke, ``Using {Evolutionary} {Algorithms} to {Target} {Complexity} {Levels} in {Game} {Economies},'' \emph{IEEE Trans. on Games}, vol.~15, no.~1, pp. 56--66, 2023.

\bibitem{Roohi2019}
S.~Roohi and A.~Forouzandeh, ``Regarding color psychology principles in adventure games to enhance the sense of immersion,'' \emph{Entertainment Computing}, vol.~30, p. 100298, 2019.

\bibitem{rupp2024balance}
F.~Rupp, A.~Puddu, C.~Becker-Asano, and K.~Eckert, ``{It might be balanced, but is it actually good? An Empirical Evaluation of Game Level Balancing},'' in \emph{2024 IEEE Conference on Games (CoG)}, 2024.

\bibitem{stiegler2018SymbolicReasoningHearthstonea}
A.~Stiegler, K.~P. Dahal, J.~Maucher, and D.~Livingstone, ``Symbolic {{Reasoning}} for {{Hearthstone}},'' \emph{IEEE Transactions on Games}, vol.~10, no.~2, pp. 113--127, Jun. 2018.

\bibitem{Tractinsky2014}
N.~Tractinsky, ``Visual aesthetics,'' in \emph{The Encyclopedia of Human-Computer Interaction}.\hskip 1em plus 0.5em minus 0.4em\relax Interaction Design Foundation, 2014.

\bibitem{wellerdiek2015strong}
A.~C. Wellerdiek, M.~Breidt, M.~N. Geuss, S.~Streuber, U.~Kloos, M.~J. Black, and B.~J. Mohler, ``Perception of strength and power of realistic male characters,'' in \emph{Proceedings of the ACM SIGGRAPH Symposium on Applied Perception}.\hskip 1em plus 0.5em minus 0.4em\relax ACM, 9 2015, pp. 7--14.

\bibitem{xiao2023MasteringStrategyCard}
C.~Xiao, Y.~Zhang, X.~Huang, Q.~Huang, J.~Chen, and c.~Sun, ``Mastering {{Strategy Card Game}} ({{Hearthstone}}) with {{Improved Techniques}},'' in \emph{2023 {{IEEE Conference}} on {{Games}} ({{CoG}})}, Aug. 2023, pp. 1--8.

\end{thebibliography}

\end{document}